\documentclass[preprint,12pt,authoryear]{elsarticle}

\usepackage{float}
\usepackage{algorithm}
\usepackage{algorithmic}
\usepackage{amsmath}
\usepackage{hyperref}
\usepackage[utf8]{inputenc}

\newtheorem{definition}{Definition}

\usepackage{amssymb}
\journal{European Journal of Operational Research}

\biboptions{authoryear}

\begin{document}

\begin{frontmatter}



\title{Many-Objective Pareto Local Search}


\author{Andrzej Jaszkiewicz}

\address{Poznan University of Technology, Faculty of Computing, Institute of Computing Science, ul. Piotrowo 2, 60-965 Poznan, andrzej.jaszkiewicz@put.poznan.pl}

\begin{abstract}
We propose a new Pareto Local Search Algorithm for the many-objective combinatorial optimization. Pareto Local Search proved to be a very effective tool in the case of the bi-objective combinatorial optimization and it was used in a number of the state-of-the-art algorithms for problems of this kind. On the other hand, the standard Pareto Local Search algorithm becomes very inefficient for problems with more than two objectives. We build an effective Many-Objective Pareto Local Search algorithm using three new mechanisms: the efficient update of large Pareto archives with ND-Tree data structure, a new mechanism for the selection of the promising solutions for the neighborhood exploration, and a partial exploration of the neighborhoods. We apply the proposed algorithm to the instances of two different problems, i.e. the traveling salesperson problem and the traveling salesperson problem with profits with up to 5 objectives showing high effectiveness of the proposed algorithm.
\end{abstract}


\end{frontmatter}

%\linenumbers

\section{Introduction}
\label{S:1}

Combinatorial optimization problems find numerous applications in transportation, logistics, scheduling, design, etc. \citep{yu2013industrial}. On the other hand, real-life optimization problems often require taking into account multiple points of view corresponding to multiple objectives. Since, multiple objective combinatorial optimization problems are usually NP-Hard, multiobjective metaheuristics are often used to solve them approximately \citep{Talbi2012}. 

Effective single objective algorithms for combinatorial optimization problems usually use some kind of local search as one of their components. Pareto Local Search (PLS) ~\citep{Angel04,Paquete2006, Paquete2007} is a very natural extension of the concept of the local search to the multiobjective case. PLS works with a Pareto archive, i.e. a set of potentially Pareto-optimal solutions that are not dominated by any other solution obtained so far. PLS explores neighborhood of each solution in the archive and uses the neighbor solutions to update the Pareto archive.

Alike the single objective local search, PLS when used as a standalone algorithm \emph{started with an initial archive of a low quality} is not very effective, however, it is an important component of a number of the state-of-the-art hybrid algorithms e.g. for the bi-objective knapsack \citep{Lust12}, the bi-objective traveling salesperson problem (TSP) \citep{Lust10b,Lust10,Liangjun14,Jaszkiewicz2017,CORNU2017314}, various bi-objective permutation flowshop problems \citep{DuboisLacoste2011}, the bi-objective set covering problem~\citep{Lust14}, and the risk-cost optimization for procurement planning \citep{MORI201788}. 

The effectiveness of PLS as a component of the hybrid algorithms may be explained by the fact that it has different characteristics than most other multiobjective metaheuristics. As noticed by ~\cite{Lara10} each multiobjective metaheuristic should search both towards and along the Pareto front. The search towards the Pareto front means generating new solutions lying closer to the Pareto front than known solutions, and the search along the Pareto front means generating new potentially Pareto-optimal solutions improving the representation of the Pareto front. PLS is very effective in the search along Pareto front when started from a seed of high quality solutions. For example in the case of the bi-objective traveling salesperson problem such a seed may be generated with the use of the Lin-Kernighan~\citep{Lin73} heuristic \citep{Lust10b,Lust10,Liangjun14,CORNU2017314}.

The standard PLS algorithm becomes, however, very inefficient in the case of more than two objectives because the number of Pareto-optimal solutions grows very fast with the number of objectives. Thus PLS has to search neighborhoods of a huge number of solutions and the process of updating large Pareto archives with new solutions becomes very time-consuming. Indeed, with very few exceptions (e.g. \cite{Liangjun14,CORNU2017314}), majority of the publications on PLS concerns bi-objective optimization only. 

In this paper we propose a new Many-objective PLS algorithm (MPLS) dedicated to the problems with more than two objectives. Please note, that we use the term "Many-objective" in somehow non-standard way, since it usually describes problems with $\geq$ 4 objectives \citep{CHAND201535}. However, in the context of PLS, three objectives is already a high number.

The proposed algorithm is based on the following three new mechanisms that differ it from the standard PLS:
\begin{itemize}
\item The use of a recently proposed ND-Tree data structure \citep{Jaszkiewicz2016arXiv160304798J} for the efficient update of the Pareto archive. 
\item The use of a new mechanism for the selection of the promising solutions for the exploration of their neighborhoods based on the randomly selected  weighted Chebycheff scalarizing functions.
%. In our algorithm we draw at random a weight vector defining a weighted Chebycheff scalarizing function and then we select from the archive solution being the best for this function. We again use ND-Tree data structure to efficiently identify the best solution for a given weighted Chebycheff scalarizing function which is a novel application of this data structure introduced in this paper. 
\item The partial exploration of the neighborhoods, i.e. testing only some of the neighborhood solutions.

\end{itemize}

The first mechanism reduces the time needed to update the Pareto archive, while the two latter mechanisms improve the quality of the archive with a limited number of the neighborhood solutions tested.

We apply MPLS to two different multiobjective combinatorial optimization problems: the multiobjective traveling salesperson problem (MTSP) and the multiobjective traveling salesperson problem with profits (MTSPWP) with up to 5 objectives. Through a computational study we show that MPLS produces archives of much better quality than the standard PLS in the same time. We also show that each of the three new mechanisms described above contributes to the effectiveness of the proposed method. Furthermore, since for the bi-objective traveling salesperson problem the best results have been obtained by properly balancing the CPU time assigned to the Lin-Kernighan heuristic and to PLS \citep{Lust10b,Lust10,Liangjun14,Jaszkiewicz2017,CORNU2017314} we test whether a similar situation holds in the case of MTSP with more than two objectives. 

The paper is organized in the following way. In the next section we describe the proposed Many-objective Pareto Local Search algorithm. Then the computational experiment is presented. In section 4, related works are discussed. Finally, conclusions and directions for further research are presented.

\section{Many-objective Pareto Local Search algorithm}
We start the description of the proposed MPLS algorithm with a short presentation of the standard PLS that is a basis for our algorithm. Then we describe the three new mechanisms used in MPLS. Finally, we present the whole algorithm of MPLS.

\subsection{Pareto Local Search}

PLS is a relatively straightforward adaptation of the idea of the single objective local search to the multiobjective case. In PLS the acceptance of a new neighborhood solution is based on the dominance relation. 

Consider a general \emph {multiobjective optimization problem} with a feasible set $\mathcal{X}$ and $d$ objective functions $y_k(x)$. Without loss of generality we assume that the objectives are maximized. The image of a feasible solution $x$ in the objective space $\mathbb{R}^d$ is a point $y(x)= (y_1(x),y_2(x),\ldots,y_d(x))$. We say that a point $y^1 \in \mathbb{R}^d$ \emph{dominates} a point $y^2 \in \mathbb{R}^d$ if, and only if, $y^1_j \geq y^2_j\  \forall \, j \in \{1,\ldots,d\} \wedge \exists \, j \in \{1,\ldots,d\}: y^1_j > y^2_j$. We denote this relation by $y^1 \succ y^2$. For the sake of simplicity we will use the dominance relation with respect to the corresponding solutions as well. The solutions not dominated by any other feasible solution are called \emph{Pareto-optimal} and the image of the set of Pareto-optimal solutions in the objective space is the \emph{Pareto front}.

A set of distinct and mutually non-dominated solutions, i.e. solutions such that none of them dominates any other, is called a \emph{Pareto archive}. In the context of multiobjective metaheuristics the Pareto archive is used to store the set of potentially Pareto-optimal solutions generated by the method till a given iteration.

PLS works with a Pareto archive. It starts with some initial Pareto archive and then searches whole neighborhoods of all solutions whose neighborhoods have not been searched yet for new potentially Pareto-optimal solutions until no such new solution can be found
in any of the neighborhoods (see Algorithm \ref{algoPLS} \emph{based on the particular version of PLS used e.g. in \cite{Lust10b}}).

\floatname{algorithm}{Algorithm}
\begin{algorithm}[!h]
\caption{\texttt{PLS}}\label{algoPLS}
\begin{algorithmic}%[1]
\STATE Parameter $\updownarrow$: $\mathcal{A}$: an initial Pareto archive
\vspace*{1\baselineskip}
\STATE $P \leftarrow \mathcal{A}$
\STATE $P_a \leftarrow \emptyset$
\WHILE{$P \neq \emptyset$}  
\FORALL{$x \in P$}  
\FORALL{$x' \in \mathcal{N}(x)$}
\IF{$y(x) \nsucc y(x')$ and $y(x) \neq y(x')$}
\IF{\texttt{Update}($\mathcal{A} \updownarrow $,$x' \downarrow$)}
\STATE \texttt{$P_a \leftarrow P_a \cup \{x'\}$}
\ENDIF
\ENDIF
\ENDFOR
\ENDFOR
\STATE $P \leftarrow P_a$
\STATE $P_a \leftarrow \emptyset$
\ENDWHILE
\vspace*{1\baselineskip}
\STATE where $\mathcal{N}(x)$ denotes the neighborhood of $x$ and \texttt{Update()} updates the Pareto archive $\mathcal{A}$.
\end {algorithmic}
\end{algorithm}

\subsection{Mechanism I: Efficient update of the Pareto archive with the use of ND-Tree}

Whenever a new neighbor solution not dominated by the current solution is found, the Pareto archive needs to be updated. Updating a Pareto archive $\mathcal{A}$ with a new solution $x'$ means that:
\begin{itemize}
\item $x'$ is added to $\mathcal{A}$ if it is not dominated nor equal to any solution in $\mathcal{A}$,
\item all solutions dominated by $x'$ are removed from $\mathcal{A}$.
\end{itemize}

Since the neighborhood solutions for many combinatorial problems may be generated in a very short time and the number of Pareto-optimal solutions grows fast with growing number of objectives, the process of updating the Pareto archive may easily become the main factor influencing the running time of PLS. 
%For example, the typically used 2-edge exchange move for the traveling salesperson problem requires just 4 arithmetic operations for each objective to generate a neighbor solution while the update of the Pareto archive with a naive approach may require thousands or millions of comparisons of solutions.

The simplest approach to the update of the Pareto archive is to organize it as a simple list of solutions and compare the new solution to each solution in this list until a dominating or equal solution has been found or all solutions have been compared. This approach is, however, very time consuming for large Pareto archives. In the bi-objective case the Pareto archive may be efficiently updated using the binary search in an archive sorted according to the values of the objectives \citep{Jaszkiewicz2016arXiv160304798J}. This approach, however, cannot be extended to the case of more than two objectives.

In MPLS we use a recently proposed ND-Tree data structure \citep{Jaszkiewicz2016arXiv160304798J}. In ND-Tree the Pareto archive is recursively split into disjoint subsets. For each subset approximate local ideal and approximate local nadir points, i.e. points in the objective space respectively dominating and dominated by all solutions in this subset, are kept. In other words, all solutions from each subset are contained in the hyper-box defined by the two points. These hyper-boxes may in general overlap, but the tree is built such that the subsets contain close solutions contained in small hyper-boxes. By comparing the new solution to the approximate local ideal and the approximate local nadir points many branches of the tree can be omitted. The computational complexity of updating the Pareto archive with ND-Tree is sub-linear with respect to the size of the archive for any number of objectives under mild assumptions \citep{Jaszkiewicz2016arXiv160304798J}. Thus the overall running time of MPLS can be significantly reduced with the use of this data structure.

\subsection{Mechanism II: Selection of the promising solutions for the exploration of their neighborhoods}

The standard PLS algorithm ends-up with a Pareto-archive being locally optimal, i.e. when no solution being a neighbor of any solution in the archive can improve this archive. Obtaining a locally optimal Pareto-archive is, however, usually prohibitively time-consuming in the case of three and more objectives even with the use of ND-Tree. On the other hand, the algorithm when stopped earlier may produce an archive of a low quality, because there could be high variations in the CPU times spent in exploring various regions of the Pareto front. \cite{DuboisLacoste2011,DUBOISLACOSTE2015} proposed anytime PLS algorithm for the bi-objective case whose goal is to generate archives of a good quality at any iteration. One of the main mechanisms of that algorithm is the selection of solutions whose neighborhood exploration has a high potential to improve the current archive. Dubois-Lacoste et al.
use the so-called optimistic hypervolume to select such solutions in the bi-objective case. This mechanism, however, cannot be directly extended to the case of more than two objectives. Thus, we propose a new selection mechanism described below.

In each iteration we would like to generate a new solution highly improving the quality of the Pareto archive. Since the neighbor solutions are usually similar to the original solutions, it is natural to expect that new good solutions may be found in the neighborhoods of known good solutions. 

Assume that an indicator of the quality of the Pareto archive is used. A good solution is a solution that substantially contributes to the value of this indicator, i.e. its removal would substantially deteriorate the value of this indicator. A number of indicators for the evaluation of the quality of Pareto archives have been proposed. An often used quality indicator is the hypervolume of the space dominated by the set of points (see e.g. \citep{Zitzler2003}). Our approach is motivated by another quality indicator - the expected value of the weighted Chebycheff scalarizing functions \citep{Jaszkiewicz02a}. Such functions are defined in the following way:
\begin{displaymath}
s_\infty(y(x), y^0,\Lambda) = \underset {k=1,...,d} \max \{ \lambda_k \, (y_k^0 - y_k(x)) \}
\end{displaymath}
where $y^0$ is a reference point, $\Lambda=[\lambda_1,...,\lambda_d]$ is a weight vector such that $\lambda_k \geq 0 \; \forall k$. Each weighted Chebycheff scalarizing function has at least one global minimum belonging to the set of Pareto-optimal solutions. For each Pareto-optimal solution $x$ there exists a weighted Chebycheff scalarizing function $s_\infty$ such that $x$ is a global minimum of $s_\infty$ (see~\citep{Steuer86}, ch. 14.8). 

The quality indicator is defined as:

\begin{displaymath}
R (\mathcal{A}, \mathcal{L}) =
\mathbb{E}_{\Lambda \in \mathcal{L}} (\underset {y \in \mathcal{A}} \min s_\infty(y, y^0,\Lambda))
\end{displaymath}
where $\mathcal{L}$ is a set of all weight vectors.

\begin{figure}[h]
\centering
\includegraphics[width=0.8\textwidth]{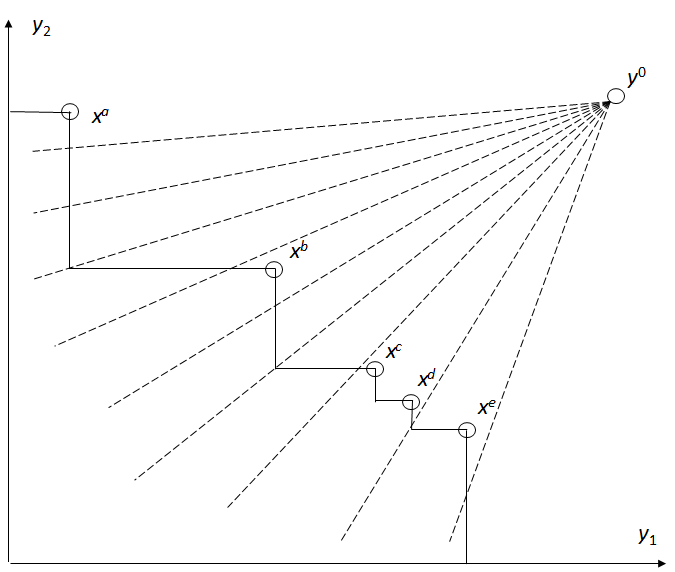}
\caption{Comparison of the hypervolume and the expected value of the weighted Chebycheff scalarizing functions indicators}
\label{fig:indicators}
\end{figure}

Although the two indicators, i.e. the hypervolume and the expected value of the weighted Chebycheff scalarizing functions, are defined in quite different terms, they are in fact closely related, which is illustrated in Figure \ref{fig:indicators}. Consider the non-dominated border of the dominated area used in the hypervolume indicator. Each point lying on this border is optimum of one or more weighted Chebycheff scalarizing functions and has the same values of these scalarizing functions as some of the solutions in the archive. In other words, the hyper-facets of this border belong to the same isoquants of multiple Chebycheff functions as some solutions from the archive. In result, any improvement of the hypervolume implies an improvement of the best values of some weighted Chebycheff scalarizing functions.

An important advantage of the expected value of the weighted Chebycheff scalarizing functions indicator is that to select good solutions we do not need to explicitly calculate the contribution of each solution to this quality indicator. Instead in each iteration we draw a random weight vector defining a weighted Chebycheff scalarizing function and select from the Pareto archive the solution that minimizes the value of this function. Thus, solutions being minima of many scalarizing functions, and so having high impact on the expected value, have higher chance to be selected. For example, in Figure \ref{fig:indicators} solutions $x^a$ and $x^b$ have higher chance to be selected than solutions $x^c$, $x^d$, and $x^e$.

This mechanism not only selects promising solutions but also assures a uniform coverage of all regions of the objective space. If the solutions were selected at random with uniform probability, regions that have more dense representation in the current archive would have a higher chance to be explored further. Thus, any random differences in the density would tend to be further increased, since the dense regions would be explored more than regions with a low density of solutions.

\begin{figure}[h]

\centering
\includegraphics[width=0.8\textwidth]{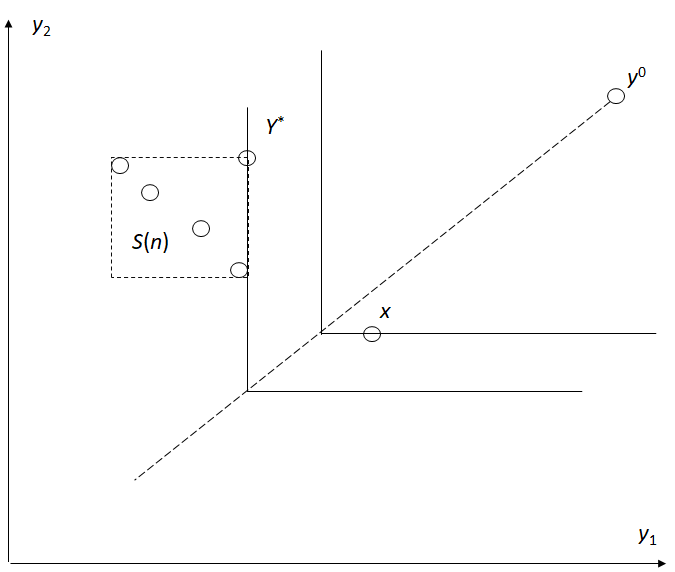}
\caption{Set $S(n)$ that could be omitted due to the knowledge of the scalarizing function value for solution $x$}
\label{fig:tree_scalarizing}
\end{figure}

Naive approach for finding the solution minimizing a given scalarizing function through evaluation of each solution would be very time consuming. However, ND-Tree data structure used to update the Pareto archive may also be used to efficiently find the solution with the best value using Algorithm \ref{algoChebMinimize} proposed in this paper. Let us remind that in ND-Tree the Pareto archive is recursively split into disjoint subsets associated with approximate local ideal points. The algorithm is based on the observation that the value of each weighted Chebycheff scalarizing function for any solution in a subset is not lower than its value for the approximate local ideal point of this subset. Thus, if a value of this function for a solution is already known (upper bound) and this value is not worse than the value for the approximate local ideal point, the whole subset could be omitted (see Figure \ref{fig:tree_scalarizing}). If this subset corresponds to an internal node, its whole subtree could be omitted. To find quickly a solution with a good value of the scalarizing function defining a good upper bound at each internal node we select first the sub-node with the best value for its approximate local ideal point.

Since Algorithm \ref{algoChebMinimize} is a new contribution reported in this paper, we analyze the computational complexity of this algorithm. This analysis is similar to the analysis of the algorithm for updating the Pareto archive with ND-Tree presented in \citep{Jaszkiewicz2016arXiv160304798J}. We start by formally defining the ND-Tree data structure :

\begin{definition}
ND-Tree data structure is a tree with the following properties:
\begin{enumerate}
\item With each node $n$ is associated a set of solutions $\mathcal{S}(n)$.
\item Each leaf node contains a list $\mathcal{L}(n)$ of solutions and $\mathcal{S}(n) = \mathcal{L}(n)$.
\item For each internal node $n$, $\mathcal{S}(n)$ is the union of disjoint sets associated with all sub-nodes of $n$.
\item Each node $n$ stores an approximate ideal point $\widehat{y^*}(\mathcal{S}(n))$ such that $\widehat{y^*}(\mathcal{S}(n)) \succeq x \quad \forall x \in \mathcal{S}(n)$ and a approximate nadir point $\widehat{y_*}(\mathcal{S}(n))$ such that $x \succeq \widehat{y_*}(\mathcal{S}(n)) \quad \forall x \in \mathcal{S}(n)$.
\item If $n'$ is a sub-node of $n$, then $\widehat{y^*}(\mathcal{S}(n)) \succeq \widehat{y^*}(\mathcal{S}(n'))$ and $\widehat{y_*}(\mathcal{S}(n')) \succeq \widehat{y_*}(\mathcal{S}(n))$.
\end{enumerate}

\end{definition}

In the worst case, at each intermediate node we need to analyze each sub-node. Thus, all solutions will need to be evaluated and the computational complexity of Algorithm \ref{algoChebMinimize} is  $\Theta(N)$ where $N$ is the number of solutions in the archive. 

In the best case, at each intermediate node only one sub-node has to be analyzed and the solutions are equally split into a predefined number of sub-nodes. In this case the number of evaluated solutions and points is described by the following recurrence:
\begin{equation}
T(N) = 2 + T(N/C) = \Theta(\log_C N)
\end{equation}
where $C$ is the number of sub-nodes. Please note, that it is possible to consider even more optimistic scenarios, e.g. when only one solution needs to be evaluated, however, the scenario analyzed above is much more realistic.

The most interesting but also the most difficult to analyze or even define is the average case. Consider the case when each internal node has exactly two sub-nodes. Assume that the number of solutions in one of the sub-nodes is drawn from the uniform distribution, i.e. each number of solutions is equally likely. The other sub-node will contain the remaining number of solutions. In an internal node, either one or both sub-nodes will be processed. Assume that the probability $p_2$ of processing both nodes is constant. In this case:

\begin{equation} \label{eq:a1}
T(N)= 2 + \frac{1}{N}(1 + p_2) \sum_{k=1}^{N-1}T(k) \\
\end{equation}
Multiply both sides by $N$:
\begin{equation} \label{eq:a2}
NT(N)= 2N + (1 + p_2)\sum_{k=1}^{N-1}T(k) \\
\end{equation}
Assume that $N \geq 2$:
\begin{equation} \label{eq:a3}
(N-1)T(N-1) = 2(N-1) + (1 + p_2)\sum_{k=1}^{N-2}T(k)
\end{equation}
Subtract equations \ref{eq:a2} and \ref{eq:a3}:
\begin{equation} \label{eq:a4}
\begin{gathered}
NT(N) - (N-1)T(N-1) = \\
2N - 2(N-1) + (1 + p_2)\big(\sum_{k=1}^{N-1}T(k) - \sum_{k=1}^{N-2}T(k)\big) \implies \\ 
T(N) = \frac{2}{N} + \frac{N+p_2}{N}T(N-1)
\end{gathered}
\end{equation}
The above recurrence has the following solution:
\begin{equation} \label{eq:a5}
T(N) = \frac{\Gamma(p_2+N+1)}{\Gamma(N+1)} - \frac{2}{p_2}
\end{equation}
\emph{where $\Gamma$ denotes the Gamma function being  an extension of the factorial to real number arguments.}
	
\noindent Since
\begin{equation} \label{eq:a6}
\lim_{N\to \infty} \frac{\Gamma(N+\alpha)}{\Gamma(N)N^\alpha} = 1
\end{equation}
$\implies$
\begin{equation} \label{eq:a7}
T(N)= \Theta(N^{p_2})
\end{equation}
The Algorithm \ref{algoChebMinimize} is sub-linear for any $p_2 < 1$ under the above assumptions. Of course, it is still a very simplified analysis of the practical behavior of the algorithm. For example, the probability $p_2$ will not, in general, be constant. In fact, it may happen that none of the sub-nodes will need to be processed, since the ideal points of all sub-nodes will have the values of the scalarizing function above the current upper bound. Such situations, would further improve the practical efficiency of the algorithm.

\floatname{algorithm}{Algorithm}
\begin{algorithm}[!ht]
\caption{\texttt{MinimizeChebycheffScalarizingFunction}}\label{algoChebMinimize}
\begin{algorithmic}%[1]
\STATE Parameter $\downarrow$: A node $n$ 
\STATE Parameter $\downarrow$: A weighted Chebycheff scalarizing function $s_\infty(\dots, \Lambda)$ 
\STATE Parameter $\updownarrow$: Minimum value of the scalarizing function known so far $s^*$
\STATE Parameter $\uparrow$: Solution $x$ minimizing the scalarizing function
\vspace*{1\baselineskip}
\IF {$n$ is a leaf node}
\STATE Calculate the value of $s_\infty(\dots, \Lambda)$ for each solution in the subset
\IF {the best value is better than $s^*$}
\STATE Return the best solution and update $s^*$
\ENDIF
\ELSE
\STATE Select the sub-node $n'$ with the minimum value of $s_\infty(\widehat{z^*}(\mathcal{S}(n')),\dots, \Lambda)$
\IF {$s_\infty(\widehat{z^*}(\mathcal{S}(n')), \dots, \Lambda) < s^*$}
\STATE \texttt{MinimizeChebycheffScalarizingFunction} ($n'$, $s_\infty(\dots, \Lambda)$, $s^*$, $x$) 
\FORALL{Remaining sub-nodes $n''$} 
\IF {$s_\infty(\widehat{z^*}(\mathcal{S}(n'')), \dots, \Lambda) < s^*$}
\STATE \texttt{MinimizeChebycheffScalarizingFunction} ($n''$, $s_\infty(\dots, \Lambda)$, $s^*$, $x$) 
\ENDIF
\ENDFOR
\ENDIF
\ENDIF

\end{algorithmic}
\end{algorithm}

\subsection{Mechanism III: Partial exploration of the neighborhoods}

In the standard PLS the whole neighborhood of each solution is explored. This strategy is not well adapted, however, to the goal of obtaining a good anytime behavior since it may lead to the situations where some regions of the Pareto front will be intensively explored while other regions will be underexplored. \cite{DuboisLacoste2011,DUBOISLACOSTE2015} and \cite{Liefooghe2012} observed that the anytime behavior of PLS can be improved by a partial exploration of the neighborhoods, i.e. testing only some neighbor solutions.

On the other hand, the algorithm may become ineffective if too few neighbor solutions are tested. Please note, that before exploring a neighborhood, the solution for the exploration needs to be selected which takes non-negligible CPU time even with the use of ND-Tree. Thus the optimum strategy may be to test a randomly selected subset of neighbor solutions. Dubois-Lacoste et al. and Liefooghe et al. consider several options for partial exploration of the neighborhoods. Our approach is slightly different since we use the number of tested moves as a parameter of the method.

\subsection{MPLS algorithm}

The proposed method is summarized in Algorithm \ref{algoMPLS}. To draw a random weight vector we use the algorithm proposed in \citep{Jaszkiewicz02a}. Weighted Chebycheff scalarizing functions are applied to the objective values normalized on-line to the range $[0, 1]$ based on the ranges of the objectives in the current archive, which is in fact achieved by dividing each individual weight by the range of the corresponding objective. 

\emph{Algorithm \ref{algoMPLS} defines the general scheme of MPLS. To apply it to a given problem a neighborhood structure has to be defined. The choice of the neighborhood structure may of course highly influence the performance of the algorithm.}

\floatname{algorithm}{Algorithm}
\begin{algorithm}[!h]
\caption{\texttt{MPLS}}\label{algoMPLS}
\begin{algorithmic}%[1]
\STATE Parameter $\downarrow$: Maximum running time $T$
\STATE Parameter $\downarrow$: Number of neighborhood moves $M$
\STATE Parameter $\updownarrow$: $\mathcal{A}$: an initial Pareto archive
\vspace*{1\baselineskip}
\WHILE{total running time is lower than $T$}  
\STATE Use as the reference point $y^0$ the point composed
of the maximum values of particular objectives in the current archive
increased by 10\% of the range of a given objective in this archive
\STATE Draw at random a weight vector $\Lambda$ defining a Chebycheff scalarizing function $s_\infty (\dots, y^0, \Lambda)$ 
\STATE Find in $\mathcal{A}$ the solution $x$ minimizing $s_\infty (\dots, y^0, \Lambda)$, i.e. call \texttt{MinimizeChebycheffScalarizingFunction} ($rn$, $s_\infty(\dots, y^0, \Lambda)$, $\infty$, $x$) with $rn$ being a root node of ND-Tree
\FOR {$M$ times}
\STATE Generate a random neighbor $x'$ of $x$
\IF{$y(x) \nsucc y(x')$ and $y(x) \neq y(x')$}
\STATE{\texttt{UpdateNDTree}($\mathcal{A} \updownarrow $,$x' \downarrow$)}
\ENDIF
\ENDFOR
\ENDWHILE

\vspace*{1\baselineskip}
\STATE where $\mathcal{N}(x)$ denotes the neighborhood of $x$ and \texttt{UpdateNDTree()} updates the Pareto archive $\mathcal{A}$ using ND-Tree data structure.
\end {algorithmic}
\end{algorithm}

\section{Computational experiment}
\label{S:exp}

To test the effectiveness of the proposed algorithm we use two different combinatorial optimization problems. \emph{All algorithms were implemented in C++ sharing as much of the code as possible. The experiment has been run on an Intel Core i7-5500U CPU at 2.4 GHz.} The source code, the instances, and the detailed numerical results are available on-line \footnote{https://sites.google.com/view/mopls/mpls}.

%in the Supplementary Materials associated with this paper.

\subsection{Multiobjective Traveling Salesperson Problem}

As one of the test problems we use the Multiobjective Traveling Salesperson Problem (MTSP) which is a typical benchmark problem for multiple objective metaheuristics.

Given a set $\{v_1, v_2, \cdots, v_N\}$ of nodes and $d$ costs $c_1(v_i,v_j) \dots c_d(v_i,v_j)$ between each pair of distinct nodes $\{v_i, v_j\}$, the multiobjective traveling salesperson problem (MTSP) consists of finding an order $\pi$ of the nodes, minimizing the following costs $(k=1,\dots,d)$:      
\begin{displaymath}
\begin{array}{l}
\textrm{``minimize''} y_{k}(\pi)=\displaystyle \sum_{i=1}^{N-1}c_k(v_{\pi(i)},v_{\pi(i+1)})+c_k(v_{\pi(N)},v_{\pi(1)})
\end{array}  
\end{displaymath}

In the computational experiment, we used the symmetric multiobjective traveling salesperson problem with $c_k(v_{i},v_{j})=c_k(v_{j},v_{i})$ for $1\leq i,j \leq N$. We used Euclidean instances with 100 and 200 nodes in which the costs of the edges correspond to the Euclidean distance between two points in a plane, and each objective corresponds to a different plane. For the purpose of this experiment, we generated 10 instances for each size and each number of objectives.

\emph{Following the successful methods proposed for the bi-objective case, we used a two-phase approach \citep{Lust10b}. In the first phase we used Lin-Kernighan heuristic~\citep{Lin73} to generate an initial Pareto archive for the second phase in which MPLS was run.} Precisely we used the efficient implementation of this heuristic from the Concorde project (http://www.math.uwaterloo.ca/tsp/concorde.html). 

We used the 2-edge exchange move in MPLS. Alike \cite{Lust10} and \cite{Jaszkiewicz2017} we used the mechanisms of candidate moves to speed-up the calculations. 

\subsection{Comparison of the various versions of PLS}
\label{S:exp1}

In the first experiment we compare MPLS to the standard PLS and to several other versions of PLS. Particular versions of PLS allow us to show not only that MPLS performs better than the standard PLS but also that each of the three new mechanisms used in MPLS indeed influences its effectiveness. The following methods are compared in this experiment:
\begin{itemize}
\item "MPLS 100" described by Algorithm \ref{algoMPLS} with 100 random moves tested in each neighborhood, which is a relatively good value of this parameter according to our preliminary experiments.
\item "MPLS full neighborhood" - the algorithm equivalent to "MPLS 100" but with the full exploration of each neighborhood. This algorithm allows us to show that the partial exploration of the neighborhoods improves the performance of MPLS.
\item "MPLS 1" described by Algorithm \ref{algoMPLS} with just one random move tested in each neighborhood. We use this algorithm to show that too small number of tested moves deteriorates the effectiveness of MPLS.
\item "MPLS 100 Random" - the algorithm equivalent to "MPLS 100" but with the random selection of solutions with uniform probability. This algorithm allows us to show that the selection of solutions based on the randomly selected weighted Chebycheff scalarizing functions improves the performance of MPLS.
\item "MPLS 100 List" - the algorithm equivalent to "MPLS 100" but with the Pareto archive organized as a simple list. This algorithm allows us to show that the use of ND-Tree data structure improves the performance of MPLS.
\item "Standard PLS tree" described by Algorithm \ref{algoPLS} but using ND-Tree data structure. This algorithm allows us to show that the selection of solutions based on the randomly selected weighted Chebycheff scalarizing functions and the partial exploration of the neighborhoods improves the performance of MPLS.
\item "Standard PLS List" described by Algorithm \ref{algoPLS}. This algorithm is the straightforward adaptation of the algorithm used in the bi-objective case to the many-objective case without any of the three new mechanisms used in MPLS. 
\end{itemize}

\emph{The number of neighborhood moves $M$ was set to 100 for all instances. It should not be considered, however, as the best value of this parameter in each case. We use this setting to proof that the best results may be achieved with some intermediate approach between testing just one move or the whole neighborhood.}

Please note, that in fact the standard PLS algorithm (see Algorithm \ref{algoPLS}) does not specify in which order set $P$ is processed. In our experiment we processed set $P$ in the natural order in which the solutions were added to this set.

\emph{In the first phase, we used 1000, 2000, and 3000 random weight vectors for $d = 3, 4, 5$, respectively. For each of the weight vectors the Lin-Kernighan heuristic was run optimizing a weighted sum of the objectives. Then, in the second phase MPLS/PLS was allowed to run for the maximum time equal to the running time of the first phase.}

\begin{figure}[h]
\centering
\includegraphics[width=0.9\textwidth]{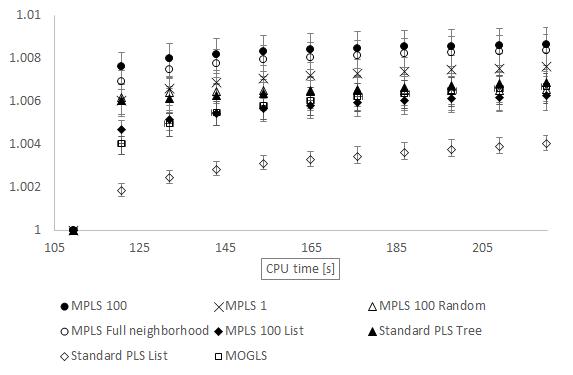}
\caption{Hypervolume indicator for 3-objective MTSP instances with 100 nodes}
\label{fig:V3}
\end{figure}

\begin{figure}[h]
\centering
\includegraphics[width=0.9\textwidth]{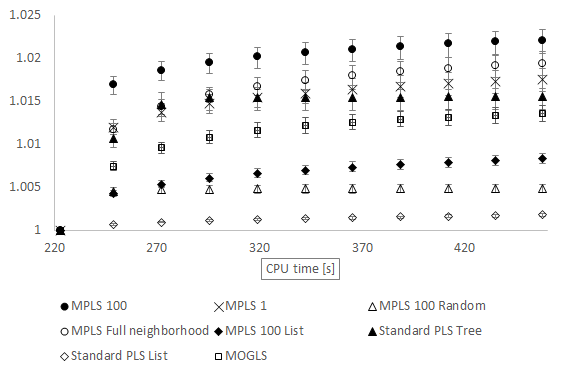}
\caption{Hypervolume indicator for 4-objective MTSP instances with 100 nodes}
\label{fig:V4}
\end{figure}

\begin{figure}[h]
\centering
\includegraphics[width=0.9\textwidth]{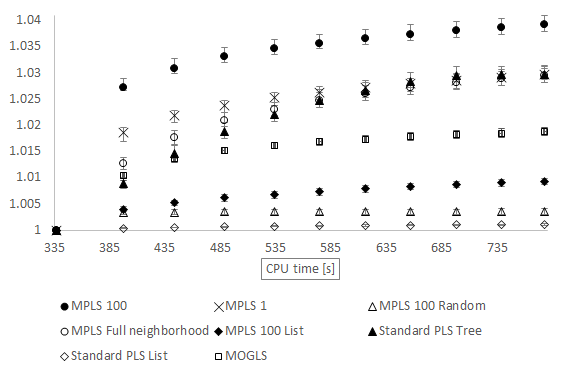}
\caption{Hypervolume indicator for 5-objective MTSP instances with 100 nodes}
\label{fig:V5}
\end{figure}

\begin{figure}[h]
\centering
\includegraphics[width=0.9\textwidth]{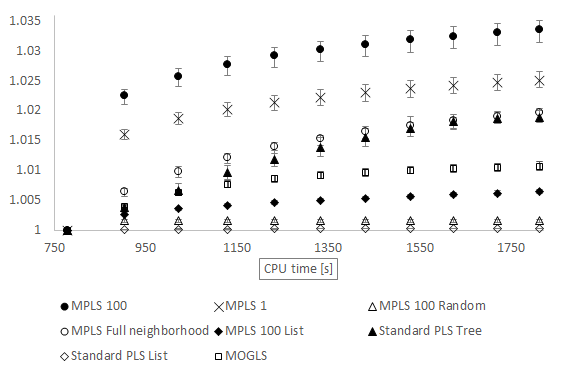}
\caption{Hypervolume indicator for 5-objective MTSP instances with 200 nodes}
\label{fig:V5_200}
\end{figure}

The results of this experiment are presented in Figures \ref{fig:V3} to \ref{fig:V5_200}. \emph{For each size and each number of objectives 10 different instances were used. As the quality measure we report the hypervolume indicator, but we calculated also the $R$ indicator, i.e. average value of the weighted Chebycheff scalarizing functions \cite{Jaszkiewicz02a} (i.e. approximation of the expected value) with exactly the same conclusions. The detailed numerical results of the two quality indicators are available on-line \footnote{https://sites.google.com/view/mopls/mpls}. Each point represents the average value and the range of the hypervolume indicator on 10 different instances obtained by a given method. The first point always corresponds to the Pareto archive obtained after the first phase. Since we used different instances we normalize the values of the hypervolume such that it is equal to 1 after the first phase.}

\emph{The reference points for the hypervolume were set to approximate nadir points of each instance with objective values multiplied by 1.5. To obtain an approximate nadir point for a given instance, each objective was optimized individually with either Lin-Kernighan heuristic (MTSP) or local search (MTSPWP). Then the worst obtained values of each objective were used as the coordinates of the approximate nadir point.}

\emph{The reported running times include also the running times of the first phase. For each instance, the maximum running time in the second phase was equal to the running time of the first phase. The values of the quality indicators were calculated in time steps equal to 10\% of the running time of the first phase. The running times were averaged over 10 runs in 10 instances. Since the variations of the running times of the first phase were relatively low, we report just the average times.}

The main observations are:
\begin{itemize}
\item The relative differences between the methods grow with the growing number of objectives. In other words, the more objectives the more important are the three new mechanisms used in MPLS.
\item Each of the three new mechanisms of MPLS is important for the final performance of the method. The comparison of "MPLS 100" to "MPLS 100 List" indicates that the mechanism with the highest influence on the effectiveness of MPLS is the use of ND-Tree data structure to update the Pareto archive and to select the best solution for a given scalarizing function. The comparison of "MPLS 100" to "Standard PLS Tree" indicates that the second mechanism with the highest influence on the effectiveness of MPLS is the selection of solutions based on the randomly selected weighted Chebycheff scalarizing functions. Furthermore, the comparison to "MPLS 100 Random" indicates that the random selection of solutions with uniform probability is even worse that the selection mechanism used in the standard PLS. The comparison of "MPLS 100" to "MPLS Full neighborhood" and "MPLS 1" indicates that the partial exploration of the neighborhoods also substantially improves the performance of MPLS.
\item"Standard PLS List", i.e. the algorithm that uses none of the three new mechanisms used in MPLS is the worst algorithm in all cases.
\end{itemize}

In order to test the significance of the differences we used the non-parametric statistical test of Mann-Whitney~\citep{Ferguson67}. The level of risk of the test has been fixed to $5\%$. The final results of "MPLS 100" were significantly better than the final results of all other methods in all cases except of "MPLS full neighborhood" with $d=3$.

\emph{We have made some additional experiments with full MPLS using $1_\nsucc$ and $1_\prec$ stopping conditions described in \cite{Liefooghe2012}. With these stopping conditions the exploration of a neighborhood was stopped after finding the first solution non-dominated with respect to the current one ($1_\nsucc$) or the first solution dominating the current one ($1_\prec$). These results are available on-line (see Section \ref{S:exp}) but not included in this paper since they were very similar to other algorithms tested. $1_\prec$ version behaved very similarly to ”MPLS full neighborhood”. This is because unlike Liefooghe et al. (2012) we initialize the archive with very high quality solutions. In result we observed that only very rarely a dominating solution was found in a neighborhood. Thus, in most cases the whole neighborhood was tested as in ”MPLS full neighborhood”. On the other hand $1_\nsucc$ stopping condition resulted in testing very few moves (often just one) since a solution non-dominated with respect to the current one was quickly found. Again please note, that unlike Liefooghe et al. (2012) we use the concept of candidate moves that increases the chance of testing high quality moves only.}

\emph{In addition, we compared MPLS to the Multiobjective Genetic Local Search (MOGLS) algorithm with path relinking \citep{Jaszkiewicz2009} for the multiobjective traveling salesperson problem. According to our knowledge the only methods reported in the literature that improved the results of this version of MOGLS are those using PLS. In this paper, we used a newer, in fact simplified, version of MOGLS in which the solutions for recombination are selected directly from the Pareto archive without any additional population of solution \citep{AghabeigJ17}. The initial archive of MOGLS was exactly the same as in the case of MPLS. The expected rank of recombined solutions was set to 20 based on preliminary experiments. MOGLS outperformed several variants of PLS/MPLS including ”Standard PLS List” and ”MPLS 100 Random” but was significantly outperformed by the full version of MPLS, i.e. "MPLS 100".}

\subsection{MPLS with various number of runs of the Lin-Kernighan heuristic}

In the above experiment we have shown that MPLS performs better than other versions of PLS. \emph{The question remains, however, whether it is competitive to Lin-Kernighan heuristic standalone applied alike in the initial phase.} Since, in the bi-objective case the best results have been obtained combining the phases of the Lin-Kernighan heuristic and of PLS \citep{Lust10b,Lust10,Liangjun14,Jaszkiewicz2017,CORNU2017314}, we test whether MPLS can improve the results obtained with Lin-Kernighan heuristic standalone also in the many-objective case. Precisely, we ask the following question: Having some predefined CPU time, can we obtain the best result with the use of the Lin-Kernighan heuristic standalone, or by combining it with MPLS in two phases.

In the first phase, we used 2000, 4000, and 6000 as the maximum number of random weight vectors for $d = 3, 4, 5$, respectively. For each of the weight vectors the Lin-Kernighan heuristic was run optimizing a weighted sum of the objectives. The time needed for all runs of the Lin-Kernighan heuristic was then used as the maximum running time. Then, the number of Lin-Kernighan heuristic runs was reduced to 90\%, \dots, 10\% and the remaining running time was allocated to MPLS. \emph{The numbers of weight vectors and thus the numbers of Lin-Kernighan heuristic runs were twice higher than in the previous experiment (see Section \ref{S:exp1}). Please note, however, that in the first experiment the same CPU time as in the fist phase was then allocated to MPLS/PLS. In the second experiment the number of weight vectors was gradually reduced and only the remaining time was allocated to MPLS/PLS. Thus, the total running time was approximately the same in both experiments.}

\begin{figure}[h]
\centering
\includegraphics[width=0.9\textwidth]{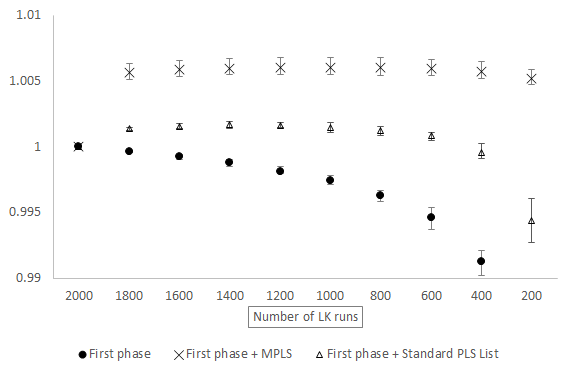}
\caption{Hypervolume indicator for 3-objective MTSP instances with 100 nodes with various numbers of the Lin-Kernighan heuristic runs and constant total CPU time}
\label{fig:B3}
\end{figure}

\begin{figure}[h]
\centering
\includegraphics[width=0.9\textwidth]{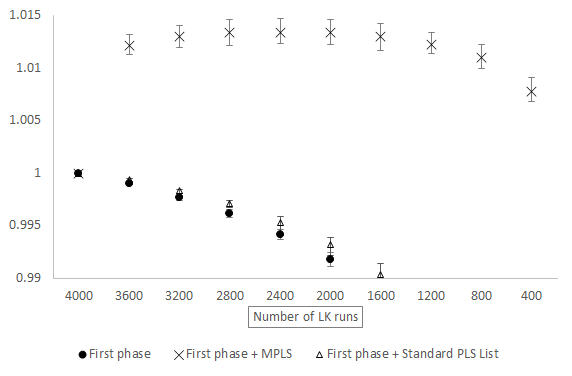}
\caption{Hypervolume indicator for 4-objective MTSP instances with 100 nodes with various numbers of the Lin-Kernighan heuristic runs and constant total CPU time}
\label{fig:B4}
\end{figure}

\begin{figure}[h]
\centering
\includegraphics[width=0.9\textwidth]{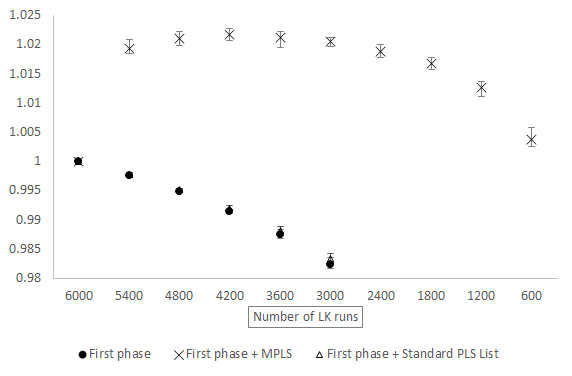}
\caption{Hypervolume indicator for 5-objective MTSP instances with 100 nodes with various numbers of the Lin-Kernighan heuristic runs and constant total CPU time}
\label{fig:B5}
\end{figure}

\begin{figure}[h]
\centering
\includegraphics[width=0.9\textwidth]{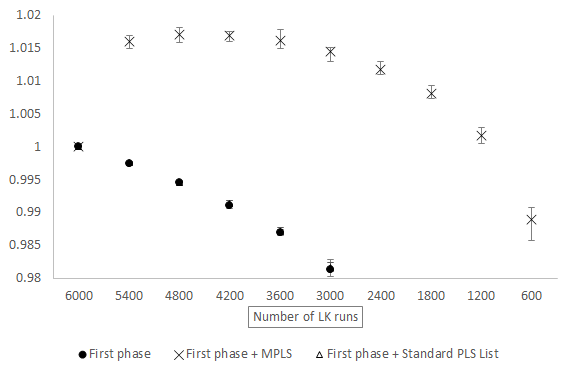}
\caption{Hypervolume indicator for 5-objective MTSP instances with 200 nodes with various numbers of the Lin-Kernighan heuristic runs and constant total CPU time}
\label{fig:B5_200}
\end{figure}

The results are presented in Figures \ref{fig:B3} to \ref{fig:B5_200}. The values of the hypervolume are again normalized such that it is 1 after the first phase with the maximum number of the Lin-Kernighan heuristic runs. Please note, that for the lower numbers of the Lin-Kernighan heuristic runs the quality after the first phase is not shown since it was much lower and the figures would become incomprehensible. 

The main observations are:
\begin{itemize}
\item In almost all cases, the results with the use of MPLS are better than with the use of the Lin-Kernighan heuristic standalone even with the lowest number of runs of the Lin-Kernighan heuristic. The only exception is the result for the 5-objective instances with 200 nodes with the smallest number of the Lin-Kernighan heuristic runs. This exception is caused by the fact that as observed in \cite{Jaszkiewicz2017} the number of Lin-Kernighan heuristic runs should grow with the size of an instance, while in our case, we used the same numbers for instances with 100 and 200 nodes in our experiment. 
\item The overall best results were obtained for 1000, 2400, 4200, and 4800 runs of the Lin-Kernighan heuristic, for $d=3,4,5$ with 100 nodes and $d=5$ with 200 nodes, respectively, i.e. the best results were obtained with 50\% or less of the CPU time allocated to MPLS. In other words, alike in the bi-objective case the best results are obtained by properly balancing the CPU time allocated to the phases of the Lin-Kernighan heuristic runs and MLPS. Please note, that the Lin-Kernighan heuristic, and especially its efficient implementation from the Concorde project, is an advanced method for TSP, while MPLS uses the very basic 2-edge exchange move. 
\item The standard PLS algorithm was able to improve the results obtained with the Lin-Kernighan heuristic standalone only in the 3-objective case. In the 4- and 5-objective cases, the standard PLS algorithm was not able to improve the results obtained with the Lin-Kernighan heuristic standalone for any reduction of the number of runs in the first phase. Thus, the use of this algorithm in these cases would not be beneficial.  
\end{itemize}

\subsection{Multiobjective Traveling Salesperson Problem with profits}

Given a set $\{v_1, \cdots, v_N\}$ of nodes and $d_1$ costs $c_1(v_i,v_j) \dots c_{d_1}(v_i,v_j)$ between each pair of distinct nodes $\{v_i, v_j\}$, and $d_2$ profits $pr_1(v_i) \dots pr_{d_2}(v_i)$ associated with each node $\{v_i\}$, the multiobjective traveling salesperson problem with profits (MTSPWP) consists of choosing a subset of nodes $SN$ and finding an order $\pi$ of these nodes, optimizing the following objectives:      
\begin{displaymath}
\begin{array}{l}
\textrm{``minimize''} y_{k}(\pi)=\displaystyle \sum_{i=1}^{\left\vert{SN}\right\vert-1}c_k(v_{\pi(i)},v_{\pi(i+1)})+c_k(v_{\pi(\left\vert{SN}\right\vert)},v_{\pi(1)}), k=1,\dots,d_1 \\
\textrm{``maximize''} y_{{d_1}+l}(\pi)=\displaystyle \sum_{i=1}^{\left\vert{SN}\right\vert}pr_l(v_{\pi(i)}), l=1,\dots,d_2
\end{array}  
\end{displaymath}

In other words, $d_1$ objectives correspond to the minimization of $d_1$ costs of the routes, while $d_2$ objectives correspond to the maximization of $d_2$ profits associated with the selected nodes. 

We used the symmetric costs where $c_k(v_{i},v_{j})=c_k(v_{j},v_{i})$ for $1\leq i,j \leq N$. For the purpose of this experiment we generated 10 instances with 200 nodes for each number of objectives. For $d=3$, we used one cost objective and two profit objectives, for $d=4$, we used two cost objectives and two profit objectives, and for $d=5$, we used two cost objectives and three profit objectives. The cost objectives were generated using Euclidean distances, i.e. the costs of the edges correspond to the Euclidean distance between two points in a plane, and each objective corresponds to a different plane. Profits were drawn at random with a uniform distribution from a range $[0, 2000]$. 

Please note, that despite of the similar names MTSPWP differs substantially from MTSP. In particular:
\begin{itemize}
\item The two problems have different solution spaces. In MTSPWP the solutions are defined by both a choice of the nodes, and an order of the selected nodes. In result, MTSPWP requires a different set of the neighborhood moves described below.
%\item The values of the objectives are calculated in a different way than in MTSP. This is obvious in the case of the profit objectives, but also the cost objectives are calculated in a different way than in MTSP since they depend also on the choice of the nodes.
\item MTSPWP exhibits a specific pattern of positive/negative correlations of objectives. The cost objectives are in general positively correlated because all of them improve with reducing the number of selected nodes. The profit objectives also are in general positively correlated because all of them improve with the growing number of selected nodes. On the other hand, the two groups of objectives are negatively correlated for the same reasons. 
\item The cost and profit objectives are defined by heterogeneous mathematical formulas. 
\end{itemize}

Because of the nature of MTSPWP, several types of moves were used to both select appropriate subset of nodes and optimize the route between these nodes \citep{Jozefowiez2008}:
\begin{itemize}
\item "2-edge exchange move" is the same move as in the case of MTSP. 
\item "Node delete move" removes one of the nodes from the current solution.
\item "Node insert move" inserts one of the nodes that is not selected at a given position in the route.
\item "Node exchange move" exchanges one of the nodes from the current solution for one of the nodes that is not selected. 
\end{itemize}

\emph{The design of the experiments was similar as in the case of MTSP}. In the first phase, instead of the Lin-Kernighan heuristic that cannot be applied in this case, we used the steepest local search algorithm that was testing all moves of all types before selecting the best neighbor. In MPLS the type of the move was drawn at random, and then the specific move was drawn. The steepest local search was applied to the optimization of the combined scalarizing functions being the weighted sums of the Chebycheff scalarizing function with weight $1$ and the linear scalarizing function with weight $0.5$, which was found the best option in the preliminary experiments. \emph{This version was found to perform better than the weighted sum used in the case of MTSP and the Lin-Kernighan heuristic. The steepest local search was then run for each of the randomly generated weight vectors.}

\begin{figure}[h]
\centering
\includegraphics[width=0.9\textwidth]{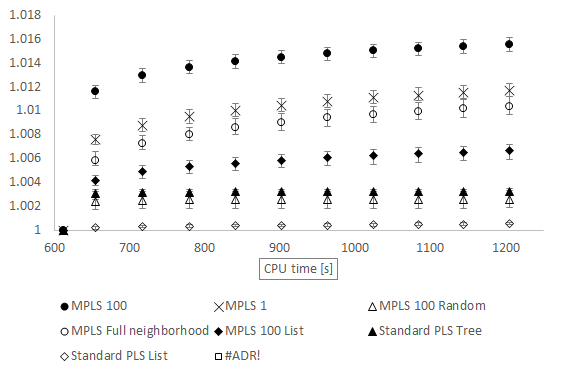}
\caption{Hypervolume indicator for 5-objective MTSPWP instances}
\label{fig:V5_TSPWP}
\end{figure}

\begin{figure}[h]
\centering
\includegraphics[width=0.9\textwidth]{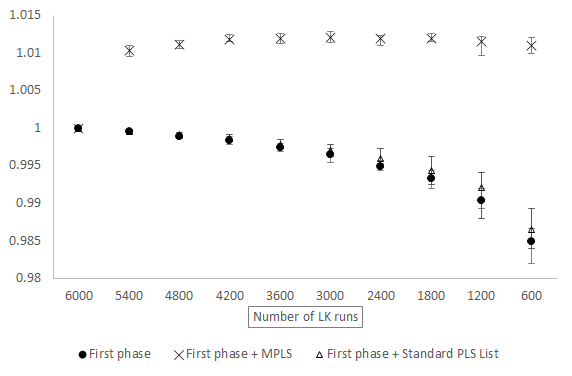}
\caption{Hypervolume indicator for 5-objective MTSPWP instances with various numbers of local search runs and constant total CPU time}
\label{fig:B5_TSPWP}
\end{figure}

We have performed the two experiments analogous like in the case of MTSP with same numbers of local search runs like the numbers of the Lin-Kernighan heuristic runs. Despite of the different nature of the two problems the results of the experiments for MTSPWP are very similar to the results for MTSP. Because of the limited space we present only the results for the 5-objective instances in Figures \ref{fig:V5_TSPWP} and \ref{fig:B5_TSPWP}. 

%The main difference with respect to the results for MTSP is that for MTSPWP the "MPLS full neighborhood" method is not significantly worse than "MPLS 100", although the average values are worse for "MPLS full neighborhood". 

\section {Related works}

Pareto Local Search algorithm has been proposed by ~\cite{Angel04}, \cite{Paquete2006}, and \cite{Paquete2007}. It has been successfully applied to the bi-objective knapsack \citep{Lust12}, the bi-objective traveling salesperson problem (TSP) \citep{Lust10b,Lust10,Liangjun14,Jaszkiewicz2017,CORNU2017314}, various bi-objective permutation flowshop problems \citep{DuboisLacoste2011}, the bi-objective set covering problem~\citep{Lust14} and the risk-cost optimization for procurement planning \citep{MORI201788}. Some recent advancements in PLS involve: Pareto improvement of some dominated solutions, i.e selection of a dominating solution from the neighborhood, until such improvements are no longer possible \citep{Inja2014}; the use of larger perturbations of solutions \citep{Drugan2012,Liangjun14}, the penalization of some features exhibited
by many solutions in the archive \citep{Alsheddy2010}, and the use of the data perturbations \citep{CORNU2017314}. However, none of these proposals focuses on the many-objective optimization.

One of the main ideas of MPLS is to improve the anytime behavior of the algorithm, since obtaining the locally optimal Pareto archive is usually prohibitively time consuming in the many-objective case. An anytime PLS algorithm has been proposed by \cite{DuboisLacoste2011,DUBOISLACOSTE2015}. Some similar concepts have been analyzed by \cite{Liefooghe2012}. Anytime PLS algorithm is, however, dedicated to the bi-objective case, and the authors define many-objective optimization as a direction for further research. At a general level some mechanisms used in MPLS are motivated by these works, but the specific techniques are very different. For example Dubois-Lacoste et al. use the optimistic hypervolume to select the promising solutions, which cannot be directly applied in the many-objective case.

\cite{Liangjun14} used PLS to solve the multiobjective knapsack problem instances with up to 4 objectives. \cite{CORNU2017314} applied PLS to the bi- and 3-objective instances of TSP. These papers do not contain, however, information whether any specific mechanisms for the many-objective case, other than the reduction of the time allocated to PLS, were used.

The idea of using the scalarizing functions with the randomly generated weight vectors has been proposed in some multiobjective evolutionary algorithms, e.g. by \cite{Ishibuchi1998}, \cite{Jaszkiewicz02a,Jaszkiewicz2009}. The scalarizing functions were, however, used in a very different way to select solutions for recombination and to optimize the scalarizing functions with the single objective local search.

The multiobjective traveling salesperson problem is often used as a benchmark problem for the multiobjective metaheuristics \citep{Jaszkiewicz02a,Garcia-Martinez2007}. As it was already mentioned, the state-of-the-art methods for the bi-objective TSP involve the use of the Lin-Kernighan heuristic and Pareto Local Search \citep{Lust10b,Lust10,Liangjun14,Jaszkiewicz2017,CORNU2017314}.

The traveling salesperson problem with profits is a well-known combinatorial optimization problem, however, it is usually considered as a single objective problem with a single cost and a single profit aggregated to one objective with a weighted sum approach \citep{Feillet2005}. There are relatively few results for the bi-objective case \citep{Jozefowiez2008,Zhu2011,Benyoucef2014,BERUBE200939}. To our knowledge, multiobjective metaheuristics have not been applied yet to the multiobjective version of the traveling salesperson problem with profits with more than one cost and/or profit.

\section{Conclusions and directions for further research}

We have presented a new Many-Objective Pareto Local Search algorithm. The algorithm has been tested on two different multiobjective combinatorial optimization problems showing its effectiveness. We have also shown that each of the three new mechanisms used in MPLS significantly contributes to its effectiveness. Furthermore, the importance of the new mechanisms in relation to the standard PLS grows with the growing number of objectives. In particular, in the 4- and 5-objective cases, standard PLS without any of the new mechanisms was not able to improve the results obtained with the Lin-Kernighan heuristic (MTSP) or local search (MTSPWP) standalone for any reduction of the number of runs in the first phase. On the other hand, MPLS was able to improve these results in almost all cases tested in the experiment.

Since the standard PLS algorithm proved its high effectiveness in the bi-objective case, we believe that MPLS could be a very useful element in the toolbox for the general case of the multiobjective combinatorial optimization, especially when combined with other methods being very effective in the search towards the Pareto front. In fact, we expect that alike the standard PLS in the bi-objective case, MPLS could become a typical component of the state-of-the-art algorithms for the many-objective combinatorial optimization.

%We believe that alike the standard PLS algorithm in the bi-objective case, MPLS could be a very useful element in the toolbox for the general case of multiobjective combinatorial optimization, especially when combined with other methods being very effective in the search towards the Pareto front. In fact, we expect that alike the standard PLS in the bi-objective case, MPLS could become a typical component of the state-of-the-art algorithms in the many-objective case.

The applications of MPLS to various benchmark and real-life problems, and combination with other multiobjective metaheuristics is a natural direction for further research. In particular MPLS could be combined with such concepts as the use of larger perturbation of solutions \citep{Liangjun14} or data perturbations \citep{CORNU2017314}.

MPLS introduces a new parameter, i.e. the number of the neighborhood moves. At present we do not have any guidelines for setting this parameter. Development of a mechanism for automatic adaptation of this parameter is an interesting direction for further research. \emph{In particular, the efficiency of the partial exploration of the neighborhood may be problem-dependent, and for some problems full exploration of the neighborhood may be still necessary.}

MPLS generates a large number of potentially Pareto-optimal solutions. The maximum size of the Pareto archive in our experiment exceeded 5 millions of solutions. The final goal of the decision maker is to select the single solution being the best according to his/her preferences. If no information about these preferences is available, a multiobjective metaheuristic should generate a good solution for any possible preferences. However, often some partial preference information is known in advance or may be gathered during the run of the method. Thus, another interesting direction for further research is the incorporation of some partial preference information in order to limit the region of the Pareto front searched by MPLS.

\section*{Acknowledgment}

This research was funded by the Polish National Science Center, grant no.~UMO-2013/11/B/ST6/01075.

%\begin{table}[h]
%\centering
%\begin{tabular}{l l l}
%\hline
%\textbf{Treatments} & \textbf{Response 1} & \textbf{Response 2}\\
%\hline
%Treatment 1 & 0.0003262 & 0.562 \\
%Treatment 2 & 0.0015681 & 0.910 \\
%Treatment 3 & 0.0009271 & 0.296 \\
%\hline
%\end{tabular}
%\caption{Table caption}
%\end{table}

%\begin{figure}[h]
%\centering\includegraphics[width=0.4\linewidth]{placeholder}
%\caption{Figure caption}
%\end{figure}



%% References


%% APA style
\bibliography{sample.bib}






% \end{thebibliography}

\end{document}